\documentclass[useAMS,usegraphicx,referee]{mn2e}

\title[Compton-thick AGN inside ULIRGs]{Compton-thick active galactic nuclei inside local ultraluminous infrared galaxies}
\author[E. Nardini \& G. Risaliti]{E.~Nardini$^{1,2}$\thanks{E-mail: nardini@arcetri.astro.it}
and G.~Risaliti$^{2,3}$ \\
$^1$ Dipartimento di Fisica e Astronomia - Sezione di Astronomia, Universit\`a di Firenze, 
L.go E. Fermi 2, 50125 Firenze, Italy\\
$^2$ INAF - Osservatorio Astrofisico di Arcetri, L.go E. Fermi 5, 50125 Firenze, Italy\\
$^3$ Harvard-Smithsonian Center for Astrophysics, 60 Garden St., Cambridge, MA 02138 USA}

\begin{document}

\date{Released Xxxx Xxxxx XX}

\pagerange{\pageref{firstpage}--\pageref{lastpage}} \pubyear{2011}

\maketitle

\label{firstpage}

\begin{abstract}
We present the X-ray analysis of the most luminous obscured active galactic nuclei (AGN) inside local ultraluminous infrared 
galaxies (ULIRGs). Our sample consists of ten sources, harbouring AGN components with estimated luminosity in excess of 
$\sim$10$^{12} L_{\sun}$ and yet unidentified at optical wavelengths because of their large obscuration. According to the 
\textit{Chandra} and \textit{XMM-Newton} spectra, only in two cases out of ten clear AGN signatures are detected at 2--10~keV 
in the shape of reflected emission. The X-ray flux from the starburst (SB) components, instead, is always broadly consistent with 
the expectations based on their IR emission. The most convincing explanation for the missing AGN detections is therefore the 
Compton-thickness of the X-ray absorber. In general, the combination of our mid-IR and X-ray spectral analysis suggests that 
the environment surrounding the AGN component in ULIRGs is much richer in gas and dust than in ordinary active galaxies, and 
the degree of AGN absorption can be tentatively related to the SB intensity, indicating a strong interaction between the two 
processes and supporting the ULIRG/quasar evolutionary scheme. 
\end{abstract}


\section{Introduction}

Accretion on to supermassive black holes (SMBHs) located in the centre of active galaxies provides a significant contribution to the 
energy irradiated over cosmic times. The spectral shape of the X-ray background and its progressive resolution into individual sources 
indicate that most active galactic nuclei (AGN) are heavily obscured by large column densities of dust and gas (e.g. Fabian \& 
Iwasawa 1999; Gilli, Comastri \& Hasinger 2007). According to the AGN unification model (Antonucci 1993), the differences observed 
in the optical spectra of AGN can be ascribed to the presence of a dusty torus surrounding the central engine: in nearly face-on objects, 
classified as type~1, favourable lines of sight allow the exploration of the very inner regions and the detection of broad emission lines on 
top of a strong optical/ultraviolet (UV) continuum. On the contrary, such spectral features are absent whenever the symmetry axis of the 
system lies close to the plane of the sky and the dusty screen blocks the direct nuclear light, so that only the high-ionization narrow 
lines originating from the outer regions are visible in type~2 objects as signatures of the underlying  accretion activity. As the dust affects 
the optical spectral properties and classification, the amount of gas along the line of sight suppresses the X-ray emission through 
photoelectric absorption and Compton scattering. Compton-thick sources (i.e. those with $N_\rmn{H} \ga 10^{24}$~cm$^{-2}$) 
are difficult to unveil at high redshift ($z \sim 1$--3) even with the deepest X-ray surveys (e.g. Alexander et al. 2003). None the less, 
the absorbed optical to soft X-ray primary radiation is reprocessed by the intervening material and re-emitted at longer wavelengths, 
driving a significant luminosity in the mid- and far-infrared (IR). Indeed, compelling evidence for the existence around $z \sim 2$ of a 
vast population of Compton-thick AGN among the IR galaxies has been found, by selecting through various criteria those sources 
showing excess mid-IR emission with respect to the predictions relative to the star formation component alone (Daddi et al. 2007; Fiore 
et al. 2008; Treister et al. 2009; Bauer et al. 2010). \\
The local counterparts of the IR systems harbouring the most obscured nuclear activity at high redshift are the so-called Ultraluminous 
infrared galaxies (ULIRGs, $L_\rmn{IR} \sim L_\rmn{bol} > 10^{12} L_{\sun}$), which rival optically-bright quasars as the most 
powerful sources in the nearby Universe with their huge mid- and far-IR emission, due to the dust reprocessing of higher-frequency 
radiation (Sanders \& Mirabel 1996). It is now well-established that the hidden source of the primary radiation field inside 
ULIRGs is a combination of extreme star formation activity (starburst, SB) and highly obscured accretion, whose clear detection has 
represented a long-standing challenge since the discovery of these objects (e.g. Genzel et al. 1998; Laurent et al. 2000). 
Therefore, two crucial points are to disentangle the AGN and SB 
components and to assess their contribution to the luminosity of both the individual sources and the local ULIRG population as a whole. 
These tasks are fundamental not only in order to understand the nature of local ULIRGs themselves, but also to shed light on the interplay 
and mutual feedback between star formation and BH accretion, which are basic ingredients of galaxy formation and evolution. \\
Taking advantage of the unprecedented data quality provided by the Infrared Spectrograph (IRS; Houck et al. 2004) onboard the 
\textit{Spitzer Space Telescope} (Werner et al. 2004) we have recently developed an AGN/SB decomposition method for local ULIRGs 
based on 5--8~$\mu$m rest-frame spectroscopy (Nardini et al. 2008). The key physical motivation that prompts the selection of this narrow 
wavelength range for our analysis is the enhancement of the intrinsic AGN over SB brightness ratio (by a factor of $\sim$10--50; Nardini 
et al. 2009, 2010) for equal bolometric luminosity, due to the intense emission from the hot dust layers exposed directly to the AGN. 
This enables the detection of even faint or heavily obscured AGN components that are missed at different wavelengths. 
Our spectral decomposition relies on a twofold observational evidence: at the highest luminosities the 5--8~$\mu$m average spectra of SB- 
and AGN-dominated sources are clearly different, and their dispersion is very limited (Brandl et al. 2006; Netzer et al. 2007). It is therefore 
possible to reproduce the AGN and SB contributions with fixed templates, only allowing for variable screen-like absorption of the AGN hot 
dust continuum occurring in the colder layers of the putative torus and/or in some dusty circumnuclear cloud.\footnote{A power-law extinction 
$\tau(\lambda) \propto \lambda^{-1.75}$ has been assumed (see also Draine 2003, and references therein).} Our main results can be summarized 
as follows: \\
1) The AGN detection rate among local ULIRGs is $\sim$70 per cent, comparable to the amount achieved by collecting all the most 
effective multiwavelength diagnostics employed so far. Remarkably, this also matches the fraction of submillimetre galaxies (Alexander et al. 2005) 
and 70 $\mu$m-selected ULIRGs (Kartaltepe et al. 2010) hosting powerful AGN activity. However, 2) the main energy supply to the bolometric 
luminosity of such objects comes from star formation, with an average AGN/SB power balance of $\sim$1/3. This ratio is confirmed to be 
luminosity-dependent, since the typical AGN contribution rises from $\sim$10 to $\sim$60 per cent across the ULIRG luminosity range. 
Consequently, the most intriguing outcome of our mid-IR analysis is 3) the identification in several sources of elusive AGN components missed 
by the standard optical diagnostic tools. In particular, we find that the sources characterized by a steep continuum (which in our model is interpreted 
as an effect of reddening), with deep absorption features and/or suppression of the polycyclic aromatic hydrocarbon (PAH) tracers of star 
formation, do not usually show the optical line ratios of active galaxies, and occasionally are classified as pure H~\textsc{ii} regions. \\
This subclass encompasses $\sim$10--15 per cent of the local ULIRG population. Once corrected for extinction, some of these elusive AGN 
components account for most of the galaxy IR luminosity, falling in the quasar -- rather than in the Seyfert -- range. As a consequence, they 
represent the ideal targets for follow-up observations in the hard X-ray band, which are relevant to the quest for the most luminous, obscured 
AGN (i.e. type 2 quasars) in the local Universe, and can also give deeper insight into the physical properties and geometrical structure of the 
obscuring material in a ULIRG-like enviroment. \\
In this work we present the X-ray analysis of a sample of ten optically-elusive and IR-obscured AGN with quasar-like luminosity, which span a 
large range with respect to the relative AGN/SB contribution, the degree of AGN obscuration and the overall IR luminosity, and are therefore broadly 
representative of their parent class. New \textit{Chandra} and \textit{XMM-Newton} observations have been obtained for four of these objects, 
i.e. IRAS~00397$-$1312, IRAS~01003$-$2238, IRAS~01298$-$0744 and IRAS~12127$-$1412, while archival X-ray spectra are available 
for the other six buried AGN completing our sample. All the sources at issue are introduced in Section~2, with the log of the X-ray observations 
and a brief description of the data reduction. In Section~3 we discuss both the existing constraints and our new results on the sources of our 
X-ray sample, combining all these indications with our previous mid-IR analysis and interpreting them in the wider context of the connection 
between circumnuclear star formation and BH growth in Section~4. The conclusions are drawn in Section~5. Throughout this paper distances 
and luminosities have been computed assuming a $\Lambda$CDM cosmology based on the latest measures of the \textit{Wilkinson Microwave 
Anisotropy Probe} ($H_0=70.5$~km~s$^{-1}$~Mpc$^{-1}$, $\Omega_\rmn{m}=0.27$ and $\Omega_\Lambda=0.73$; Hinshaw et al. 2009). 

\begin{figure}
\includegraphics[width=8.5cm]{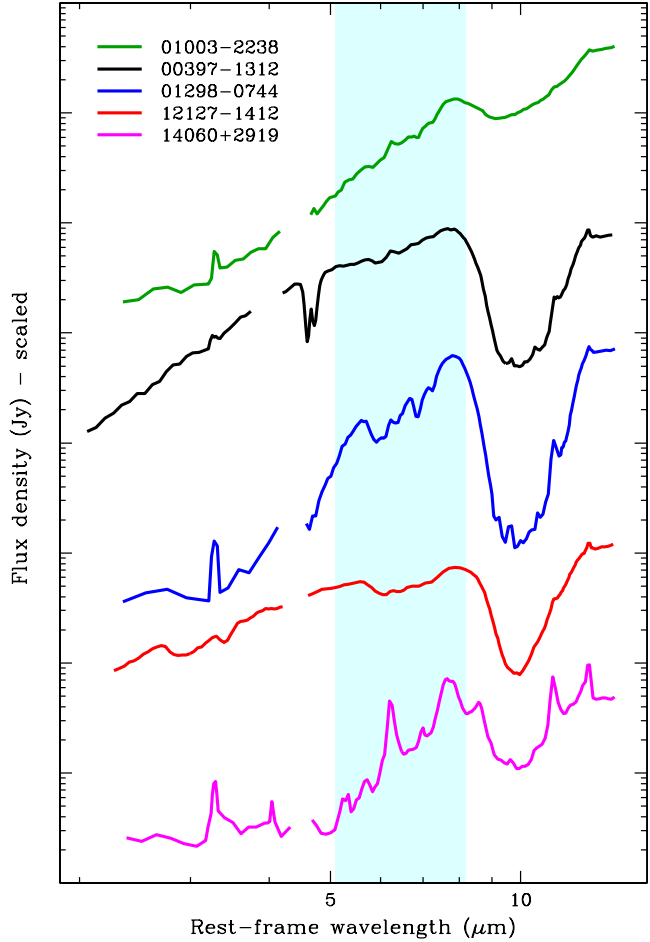}
\caption{The $\sim$2--14~$\mu$m rest-frame spectra (taken by \textit{Akari} and \textit{Spitzer}, and representative of the whole sample under study) 
of the four sources newly observed in the X-ray band. For the sake of comparison, we also show the spectrum of a typical SB-dominated ULIRG, 
IRAS~14060+2919. Each of our X-ray targets shows very faint or absent PAH emission (whose features are expected at 3.3, 6.2, 7.7, 8.6 and 
11.3~$\mu$m) and broad formless and/or narrow deep absorption (due to e.g. water ice at $\sim$3.1 and 6.0~$\mu$m, carbon monoxide at 
4.65~$\mu$m, hydrogenated amorphous carbons at 6.85 and 7.25~$\mu$m, silicates at 9.7~$\mu$m). These are all signatures of the unambiguous 
presence of a buried compact engine and are not consistent with a SB-dominated environment. The shaded region indicates the interval used for our 
AGN/SB decomposition analysis. (The \textit{Akari} spectra, courtesy of M.~Imanishi, have been rebinned for display purpose.)}
\label{f1}
\end{figure}

\begin{table*}
\caption{General properties, IR spectral parameters and X-ray observation log of the sources selected for our X-ray analysis. 
(1)~\textit{IRAS} name, (2)~redshift, (3)~IR luminosity in log of solar units, (4)~optical classification, (5)~1$\sigma$ confidence 
range for the AGN bolometric contribution (in per cent) from our mid-IR analysis, (6)~AGN optical depth at 6~$\mu$m, (7)~X-ray 
satellite and observation identifier, (8)~observation date, (9)~net exposure in ks, (10)~references.}
\label{t1}
\begin{tabular}{cccccccccc}
\hline
Object & $z$ & $L_\rmn{IR}$ & Type & $\alpha_\rmn{bol}$ & $\tau_6$ & ObsID & Date & $T$ & Ref. \\
(1) & (2) & (3) & (4) & (5) & (6) & (7) & (8) & (9) & (10) \\
\hline
00091$-$0738 & 0.118 & 12.27$\pm$0.03 & H~\textsc{ii} & 52--64 & 2.30 & \textit{Chandra}--10342 & 2008/11/01 & 15 & 4 \\
00182$-$7112 & 0.327 & 12.95$\pm$0.04 & LINER & 96--99 & $-$ & \textit{XMM}--0147570101 & 2003/04/16 & 10 & 2 \\
00397$-$1312 & 0.262 & 12.95$\pm$0.08 & H~\textsc{ii} & 50--58 & 0.24 & \textit{Chandra}--11532 & 2009/08/23 & 44 & 1 \\
01003$-$2238 & 0.118 & 12.33$\pm$0.06 & H~\textsc{ii} & 46--54 & 1.58 & \textit{XMM}--0605880101 & 2009/12/08 & 28 & 1 \\
01166$-$0844 & 0.118 & 12.12$\pm$0.06 & H~\textsc{ii} & 78--94 & 2.35 & \textit{Chandra}--10344 & 2008/10/31 & 15 & 1 \\
01298$-$0744 & 0.136 & 12.36$\pm$0.05 & H~\textsc{ii} & 71--80 & 1.79 & \textit{Chandra}--11531 & 2009/10/21 & 21 & 1 \\
07251$-$0248 & 0.088 & 12.41$\pm$0.04 & $-$ & 35--43 & 1.37 & \textit{Chandra}--7804 & 2006/12/01 & 15 & 1 \\
08572+3915 & 0.058 & 12.15$\pm$0.01 & LINER & 83--88 & 0.44 & \textit{Chandra}--6862 & 2006/01/26 & 14 & 3 \\
11095$-$0738 & 0.107 & 12.28$\pm$0.03 & LINER & 56--64 & 1.56 & \textit{Chandra}--10347 & 2009/04/09 & 16 & 4 \\
12127$-$1412 & 0.133 & 12.20$\pm$0.06 & LINER & 85--91 & $-$ & \textit{Chandra}--11533 & 2010/04/30 & 35 & 1 \\
\hline
\end{tabular}
\begin{flushleft}
\textit{References:}~$^1$This work; $^2$Nandra \& Iwasawa (2007); $^3$Teng et al. (2009); $^4$Teng \& Veilleux(2010). 
\end{flushleft}
\end{table*}

\begin{table*}
\caption{Summary of the results of our X-ray spectral analysis. }
\label{t2}
\begin{tabular}{ccccc}
\hline
Object & $F_\rmn{X}^\rmn{AGN}$ & $F_\rmn{X}^\rmn{SB}$ & $F_\rmn{X}^\rmn{obs}$ & $\epsilon_\rmn{refl}$ \\
\hline
00091$-$0738 & 5.8 & 0.023 & $<0.003$ & $<0.05$ \\
00182$-$7112 & 2.4 & 0.011 & 0.150 & 6.0 \\
00397$-$1312 & 3.3 & 0.017 & 0.007 & $<0.3$ \\
01003$-$2238 & 7.2 & 0.019 & 0.005 & $<0.2$ \\
01166$-$0844 & 4.8 & 0.015 & $<0.001$ & $<0.02$ \\
01298$-$0744 & 6.8 & 0.021 & $<0.002$ & $<0.03$ \\
07251$-$0248 & 7.4 & 0.060 & $<0.002$ & $<0.03$ \\
08572+3915 & 33 & 0.060 & $<0.005$ & $<0.02$ \\
11095$-$0738 & 8.4 & 0.028 & 0.008 & $<0.1$ \\
12127$-$1412 & 4.1 & 0.013 & 0.040 & 1.0 \\
\hline
\end{tabular}
\begin{flushleft}
\textit{Note.} $F_\rmn{X}^\rmn{AGN}$: intrinsic 2--10~keV AGN flux estimated from our mid-IR analysis, assuming standard SEDs and UV
 to X-ray corrections as described in the text; $F_\rmn{X}^\rmn{SB}$: upper limit to the 2--10~keV SB flux based on the average far-IR to X-ray 
relation for star-forming galaxies; $F_\rmn{X}^\rmn{obs}$: observed 2--10~keV flux; $\epsilon_\rmn{refl}$: rough estimate of the AGN reflection 
efficiency (in per cent). All the fluxes are in units of 10$^{-12}$~erg s$^{-1}$ cm$^{-2}$.
\end{flushleft}
\end{table*}

\section{Observations and Data Reduction}

In our previous work (Nardini et al. 2010) we have examined the 5--8~$\mu$m \textit{Spitzer}-IRS spectra of 164 local ULIRGs 
at redshift $z \la 0.35$, separating the AGN and SB components and providing a quantitative estimate of their contribution to the bolometric 
luminosity of each galaxy. This can be achieved by means of a straightforward analytical model, once the AGN emission has been corrected 
for cold dust extinction by using the reddening of the hot dust template as a measure of the optical depth (Nardini et al. 2008). It turns out that the 
optical classification is on average in good agreement with the findings of our mid-IR diagnostics. None the less, among the 20 ULIRGs 
(out of 164) for which we assess an AGN bolometric contribution ($\alpha_\rmn{bol}$) larger than 25 per cent and an optical depth to the 
AGN ($\tau_6$) larger than one at 6~$\mu$m, only three objects are optically classified as type~2 Seyferts. All the remaining entries are 
classified as either low-ionization nuclear emission-line regions (LINERs) or even H~\textsc{ii} regions, revealing that a substantial fraction 
of sources harbour significant but highly obscured nuclear activity. \\
We have therefore selected the optimal candidates for a supplementary spectral study of this population in the X-rays according to the 
following criteria, which are applied to the whole IR sample: 1)~availability of a meaningful (i.e. $>5$~ks exposure) X-ray 
observation;\footnote{The nature of the sources discussed in this work practically restricts the archival search to the present-day 
X-ray observatories, with large effective area at $\sim$2--10~keV. Even in this case, in fact, a non-detection in a very short snapshot 
would have a non-univocal interpretation.} 2)~estimated AGN luminosity exceeding $\sim$10$^{12} L_{\sun}$; 3)~non-Seyfert 
optical classification (e.g. Veilleux, Kim \& Sanders 1999). In particular, the second requirement removes any ambiguity in the AGN 
identification, since the nuclear component is always strong enough to dominate the 5--8~$\mu$m emission. The corresponding 
selection consists of six objects for which X-ray observations are already available, even if only in four cases the 
results have been published. Four additional sources, i.e. IRAS~00397$-$1312, IRAS~01003$-$2238, IRAS~01298$-$0744 and 
IRAS~12127$-$1412, all meeting the latter two criteria, have been obtained as a part of our X-ray follow-up campaign. They were 
chosen to complete the present sample, and for their remarkable expected X-ray properties: \\
\textit{(a)}~By assuming a standard quasar spectral energy distribution (SED; Elvis et al. 1994), suggesting that 
$\nu L_\nu$(2500~\AA)~$\simeq$~2$\times \nu L_\nu$(6~$\mu$m), and the most recent relations between the UV and X-ray luminosities 
(Lusso et al. 2010), the intrinsic 2--10~keV flux of the AGN component in IRAS~01003$-$2238 and IRAS~01298$-$0744 is expected to be 
$\sim 7 \times 10^{-12}$~erg s$^{-1}$ cm$^{-2}$, ensuring a prominent detection in case of a Compton-thin column density 
($N_\rmn{H} \la 5 \times 10^{23}$~cm$^{-2}$). Even in a Compton-thick prospect, allowing for a reflection efficiency of a few per cent makes 
the AGN clearly detectable with the current X-ray observatories. \\
\textit{(b)}~IRAS~00397$-$1312 and IRAS~12127$-$1412 are very interesting in this context as well (see also Imanishi et al. 2008, 2010). The former 
is optically classified as H~\textsc{ii} region in spite of being the most luminous ULIRG in the 1~Jy sample (Kim \& Sanders 1998), with 
$L_\rmn{IR} \simeq 10^{13} L_{\sun}$: its mid-IR spectrum shows faint PAH features (Fig.~\ref{f1}), and clearly hints at the simultaneous 
presence of a buried AGN roughly comparable to the SB in terms of its IR energy output. The latter is found instead near the lower end of the 
ULIRG luminosity range, but its heavily absorbed continuum lacking any star formation signature implies an AGN-dominated nature. For these 
two sources the predicted X-ray flux is just slightly less than the estimate given above (by a factor of $\sim$2). \\
The four $\sim$2--14~$\mu$m rest-frame spectra are fully representative of this ULIRG subclass, and are shown in Fig.~\ref{f1} in contrast with 
a typical SB template, while the general information concerning all the sources and their physical parameters as derived from our mid-IR analysis 
are listed in Table~\ref{t1}. Considering all the entries, this X-ray sample allows us to probe a wide range of properties with respect to obscured 
activity inside ULIRGs. It is also worth emphasizing that the 2--10~keV flux obtained by ascribing all the far-IR emission\footnote{We assume 
$F_\rmn{FIR} = 1.26 \times 10^{-11}(2.58 f_{60}+ f_{100})$~erg s$^{-1}$ cm$^{-2}$, where the \textit{IRAS} flux densities $f_{60}$ and $f_{100}$ are in 
Jy (Helou, Soifer \& Rowan-Robinson 1985).} to the SB component and applying the far-IR to X-ray correlation of Ranalli, Comastri \& Setti (2003) lies in 
the range $\sim$1--$6 \times 10^{-14}$~erg s$^{-1}$ cm$^{-2}$, which is on average $\sim$300 times\footnote{We note that our absorption correction of 
the AGN flux is $\sim$10 at most ($\tau_6 \simeq 2.35$ in IRAS~01166$-$0844), hence it does not significantly affect the contrast between the AGN and 
SB expected X-ray brightness.} lower than what envisaged for the AGN (see Table~\ref{t2}). The median 2--10~keV luminosity predicted for the AGN 
component is $L_\rmn{X} \sim 2.5 \times 10^{44}$~erg s$^{-1}$. This is actually a conservative evaluation: if we adopt the direct 6~$\mu$m to 2--10~keV 
correction found by Lutz et al. (2004a), it should be revised upwards by a factor of $\sim$2. This places our sources well above the X-ray intrinsic luminosity 
of the local population of Compton-thick AGN recently identified by Goulding et al. (2011), allowing us to explore the parameter space of highly obscured AGN 
in the low-redshift/high-luminosity region. \\
Concerning the unpublished X-ray data, IRAS~01003$-$2238 was observed by \textit{XMM-Newton} on 2009 December 08 for 35~ks. The EPIC pn and 
MOS cameras operated in full frame mode. After the filtering of high-background periods, the useful pn exposure is $\sim$28~ks. Incidentally, a previous 
snapshot ($\sim$10~ks) taken with \textit{Chandra} in 2003 yielded only 20 counts at 0.5--8.0~keV, preventing a detailed spectral modelling (Teng et al. 2005). 
The other five sources, instead, were observed with the \textit{Chandra} ACIS-S detector. The basic details about all the observations are summarized in 
Table~\ref{t1}. The data reduction was performed following the standard procedures, using the latest versions of the \textsc{ciao} and \textsc{sas} packages 
for \textit{Chandra} and \textit{XMM-Newton} event files, respectively. The source and background spectra were extracted from circular regions with radius of 
2\arcsec (\textit{Chandra}) and 30\arcsec (\textit{XMM-Newton}). In each case the X-ray emission from the source is entirely collected within the chosen 
aperture, corresponding to $\sim$3--10~kpc in the \textit{Chandra} images. No clear morphological structure can be appreciated. The analysis has 
been performed on unbinned spectra using the \textsc{xspec} v12.5 fitting package, and $C$-statistic has been adopted due to the low number of 
counts. Quoted uncertainties and upper limits are given at the 90 per cent confidence level. 

\section{Spectral Analysis and Discussion}

Since ULIRGs are in general very faint X-ray sources, only a limited number of objects in our local IR sample have been targeted  
with the current X-ray observatories \textit{Chandra} (Ptak et al. 2003; Teng et al. 2005), \textit{XMM-Newton} (Franceschini et al. 
2003) and \textit{Suzaku} (Teng et al. 2009). A compilation of all publicly available X-ray data for 40 ULIRGs has been recently 
discussed by Teng \& Veilleux (2010). \\
Analogously to their mid-IR spectra, the X-ray spectra of ULIRGs usually show the combined signatures of AGN and SB activity; in 
particular, the soft ($E < 2$~keV) emission is characterized by a diffuse thermal component associated with star formation in the 
circumnuclear regions, while at higher energies the hard AGN power law is transmitted through typical column densities of the order 
of $\sim$10$^{23}$~cm$^{-2}$. For Compton-thick sources a flat reflection spectrum is expected, alongside the prominent iron K line 
due to fluorescent emission in the surrounding optically-thick and almost neutral gas (e.g. Matt, Brandt \& Fabian 1996). In this section 
we first analyse our new observations and then review the existing X-ray spectra of the other elusive AGN in our sample, in order to 
explore the properties of the obscured AGN population within local ULIRGs and obtain some useful hints about and their high-redshift 
equivalents. 

\subsection{IRAS~00397$-$1312}

\begin{figure}
\includegraphics[width=8.5cm]{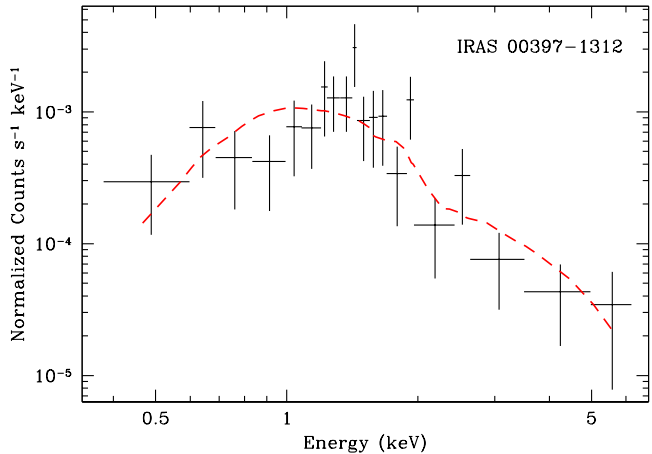}
\caption{\textit{Chandra} spectrum of IRAS~00397$-$1312 reproduced with a \textsc{tbabs*zwabs*powerlaw} model. In these figure 
and the following ones, the X-ray data are rebinned for plotting purposes.}
\label{f2}
\end{figure}

As mentioned above, IRAS~00397$-$1312 is one of the most striking objects in the local Universe for many reasons. Owing to its 
huge luminosity and composite nature, it is an ideal source for a case study of the AGN/SB connection in extreme environments. This 
ULIRG does not show clear morphological indications of a recent interaction, and likely represents an old merger stage (Veilleux, Kim \& 
Sanders 2002). Despite its optical classification as H~\textsc{ii} region, the mid-IR spectral shape reveals all the unambiguous 
signatures of a buried AGN (Imanishi 2009), first and foremost the outstanding CO absorption profile around $\sim$4.65~$\mu$m. 
The X-ray spectrum of IRAS~00397$-$1312 is shown in Fig.~\ref{f2}: the only detected X-ray emission is well reproduced by means 
of an absorbed power law, resulting in a $C$-stat of 45.0 for 52 degrees of freedom (d.o.f.). Besides the Galactic column density along the 
line of sight towards the target, which has been taken into account in all our fits, an additional X-ray absorber with 
$N_\rmn{H} \simeq 4.5 \times 10^{21}$~cm$^{-2}$ local to the source at $z=0.262$ is required. The unabsorbed 0.5--2 and 2--10~keV 
fluxes are, respectively, $\sim$9 and $7 \times 10^{-15}$~erg s$^{-1}$ cm$^{-2}$, and are both in good agreement with the SB 
X-ray flux predicted from our mid-IR analysis and the Ranalli et al. (2003) relations for star-forming galaxies (Table~\ref{t2}). The steepness 
of the power law, whose photon index is $\Gamma \simeq 2.3^{+1.1}_{-0.6}$, is itself consistent with pure SB emission with no AGN contribution; 
for the latter, an upper limit has been evaluated by adding a \textsc{pexrav} component (Magdziarz \& Zdziarski 1995) to account for possible 
AGN reflection. All the geometrical and abundance parameters have been frozen to their nominal \textsc{pexrav} values, as well as the 
photon index of the illuminating AGN power law (assumed to be 2.0). The upper limit to the observed 2--10~keV AGN flux is 
$\sim 7.6 \times 10^{-15}$~erg s$^{-1}$ cm$^{-2}$, implying a reflection efficiency $< 0.3$ per cent (Table~\ref{t2}), far below the usual values 
of $\sim$1--5 per cent. We will discuss this point further in the following. We note that the inclusion of a \textsc{mekal} component (i.e. emission 
from hot diffuse gas) improves the fit at soft energies, but makes the hard X-ray parameters badly constrained. Given the low quality of our 
data, we favour a simpler model even if the statistical significance is slightly lower.

\subsection{IRAS~01003$-$2238}

\begin{figure}
\includegraphics[width=8.5cm]{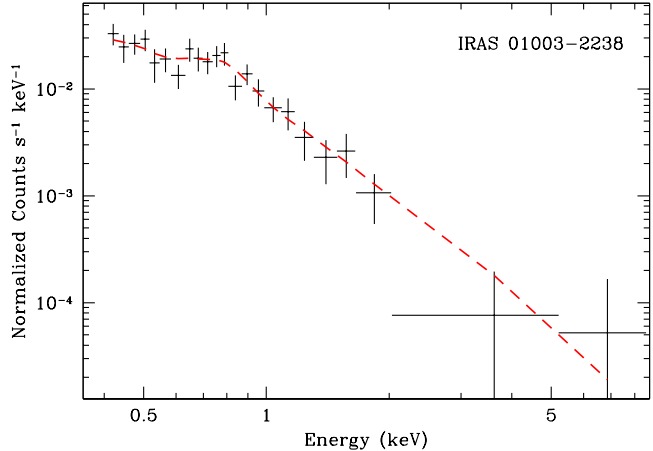}
\caption{\textit{XMM-Newton} spectrum of IRAS~01003$-$2238 and best fit obtained according to a \textsc{tbabs*(mekal+powerlaw)} 
model. Only the EPIC pn data have been plotted.}
\label{f3}
\end{figure}

This source is usually classified as H~\textsc{ii} region in the optical (Veilleux et al. 1999), yet there are also some conflicting claims 
about its possible Seyfert~2 nature (Allen et al. 1991; Yuan, Kewley \& Sanders 2010). Although detected, absorption features 
in the mid-IR are not dramatic in both depth and shape, and PAH emission is clearly distinguished on top of a very steep 
continuum. The X-ray spectrum is described with a power law plus a \textsc{mekal} component, yielding a $C$-stat/d.o.f.$= 21.3/35$. 
The spectrum is very soft (Fig.~\ref{f3}), and is characterized by a \textsc{mekal} temperature of $kT = 0.66^{+0.31}_{-0.21}$~keV and 
a power law photon index of $2.9^{+0.4}_{-0.3}$, suggesting again a sole SB origin. The observed flux is broadly consistent with such 
interpretation ($\sim 2 \times 10^{-14}$ and $5 \times 10^{-15}$~erg s$^{-1}$ cm$^{-2}$ in the 0.5--2 and 2--10~keV energy range). 
An upper limit to the 2--10~keV AGN contribution has been computed as above, resulting in $\sim 1.3 \times 10^{-14}$~erg s$^{-1}$ 
cm$^{-2}$. Hence, similar considerations can be drawn in this case: no direct AGN emission is detected below 10~keV, hinting at a 
Compton-thick circumnuclear environment. The lack of any AGN signature even in the form of a reflected spectrum implies a complete 
covering of the X-ray absorber. 

\subsection{IRAS~01298$-$0744}

This is undoubtedly the most puzzling object of the present selection, since the outcome of our X-ray observation is at 
first a striking non detection, as also confirmed with the \textsc{ciao} `wavdetect' tool. A serendipitous unidentified 
faint source is found $\sim$15\arcsec ~away from the ULIRG 
position, but any physical relation between the two can be ruled out since this angular distance is equivalent to over 
30~kpc at the redshift of IRAS~01298$-$0744, which is actually an advanced merger with only residual tidal tails. By 
considering the background count rate, we can provide an upper limit to the observed 2--10~keV source excess 
of $\sim 2 \times 10^{-15}$~erg s$^{-1}$ cm$^{-2}$, corresponding to a marginal detection with 2$\sigma$ confidence level. 
Such a limit is significantly lower than the related entries of Table~\ref{t2}, and yet reconcilable with a SB component with 
a far-IR luminosity of $\sim 2 \times 10^{11} L_{\sun}$, that would account for $\sim$10 per cent of the overall IR emission. 
Ascribing instead the upper limit to an AGN with an intrinsic X-ray flux as that predicted, the resulting column density has to 
be larger than $\sim 8 \times 10^{24}$~cm$^{-2}$. As discussed in the following, this case is not unique in its class. 

\subsection{IRAS~12127$-$1412}

\begin{figure}
\includegraphics[width=8.5cm]{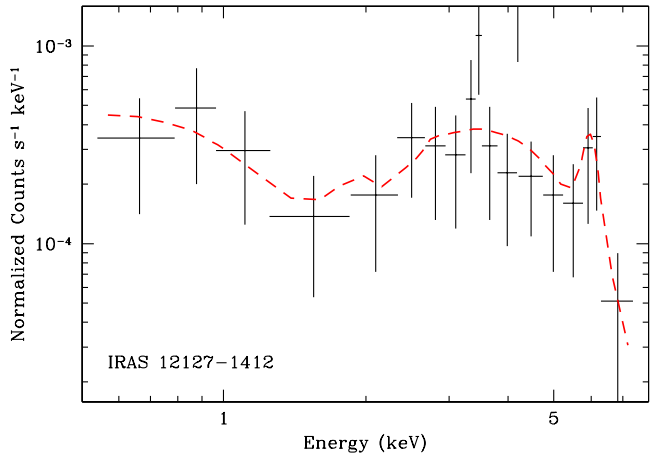}
\caption{The complex X-ray emission of IRAS~12127$-$1412, modelled as \textsc{tbabs*[powerlaw+zwabs*(zgauss+powerlaw)]} 
with $C$-stat/d.o.f.$= 48.0/47$. The observed 2--10~keV flux and the detection of a prominent iron line clearly point to AGN reflection.}
\label{f4}
\end{figure}

IRAS~12127$-$1412 is one of the few objects in our large IR sample whose 5--8~$\mu$m spectral shape cannot be 
reproduced by means of the standard AGN and SB templates. In fact, the observed emission is characterized by a 
broad formless absorption trough affecting most of the range of interest, and the putative continuum is slightly flatter than 
the assumed AGN hot dust component. This prevents us from measuring $\tau_6$ from the reddening, and yet the AGN 
obscuration is likely to be substantial: first, because a stepwise correlation between the continuum reddening and the 
presence of individual absorption features is reasonably expected (e.g. Sani et al. 2008; Nardini et al. 2009); second, 
because the steeper trend seen shortwards of 5~$\mu$m (Fig.~\ref{f1}) hints at a change in the slope of the extinction law 
(e.g. Nishiyama et al. 2009) rather than at vanishing obscuration. Based on these considerations, the intrinsic X-ray flux 
quoted in Section~2 for the AGN component of IRAS~12127$-$1412 can even be underestimated. It is therefore remarkable 
that the only clear AGN detection among our new observations unfolds in this source (Fig.~\ref{f4}). Indeed, the X-ray 
spectrum is quite complex: a steep power-law ($\Gamma \simeq 3.0 \pm 1.5$) or thermal component is needed at soft 
energies, while the hard emission can be modelled with a prominent iron feature at $\sim 6.77^{+0.50}_{-0.33}$~keV on 
top of an absorbed ($N_\rmn{H} \simeq 9^{+17}_{-5} \times 10^{22}$~cm$^{-2}$) continuum. All the parameters cannot be 
adequately constrained, but the characteristic flatness of a cold reflection spectrum is not observed, since the photon index 
of the AGN component is $\sim 2.0^{+1.1}_{-0.4}$. Also, the line energy is more consistent with He-like iron than 
with cold neutral material, pointing to ionized reflection. Unfortunately, we are not able to discuss this uncommon but intriguing 
scenario in detail. We simply note that such a possibility has already been suggested for IRAS~00182$-$7112, which is 
included in our X-ray sample as well and is remarkably similar to IRAS~12127$-$1412 both in the mid-IR and in the X-rays 
(Spoon et al. 2004; Nandra \& Iwasawa 2007), as described below. Returning to the source at issue, we estimate an 
equivalent width for the K-shell feature of $\sim$1.3~keV, with a lower limit that can be roughly placed at 0.4~keV.\footnote{A 
rigorous calculation of the line parameters is precluded by the vanishing continuum at higher energies.} The observed flux in 
the 0.5--2 and 2--10~keV range is $\sim 2 \times 10^{-15}$ and $4 \times 10^{-14}$~erg s$^{-1}$ cm$^{-2}$, and adds 
further evidence to the presence of an obscured AGN with a reflection efficiency of $\sim$1 per cent. 

\subsection{IRAS~01166$-$0844 and IRAS~07251$-$0248}

Two other objects within our selection of ULIRGs that are missed by the optical diagnostics as Seyfert galaxies but likely harbour 
an AGN with quasar-like luminosity, although observed in the X-rays, have never been published to date. We have therefore reduced 
as outlined above and then analysed the $\sim$15~ks long observations of IRAS~01166$-$0844 and IRAS~07251$-$0248 
retrieved from the \textit{Chandra} archives. The former source is a close equivalent (although much fainter) to IRAS~01298$-$0744 
at 5--8~$\mu$m; strikingly, it turns out to be completely undetected as well. The latter, instead, which is a very bright source in the \textit{IRAS} 
Revised Bright Galaxy Sample ($z \simeq 0.088$, $f_{60} \simeq 6.5$~Jy; Sanders et al. 2003) delivers a 3.5$\sigma$ detection, and its 
weak emission can be fitted with a steep power law of photon index $\simeq$3.6($\pm1.6$). In both cases, the upper limit (or estimated) 
2--10~keV flux is of the order of $\sim$10$^{-15}$~erg s$^{-1}$ cm$^{-2}$. Assuming again the far-IR to X-ray conversion of Ranalli et al. 
(2003), such a tiny value implies not only that the AGN is thoroughly absorbed, but also that the SB provides very little contribution to the 
bolometric luminosity of the host galaxy.

\subsection{The whole X-ray sample}

The remaining four sources in our sample similarly display strong evidence for buried nuclear activity at mid-IR wavelengths, with spectral 
trends that are well represented by the ones shown in Fig.~\ref{f1}. At high energies, IRAS~00091$-$0738 and IRAS~11095$-$0238 have 
been classified as \textit{weak} X-ray ULIRGs by Teng \& Veilleux (2010): the small count rate does not allow traditional spectral fitting, and 
the only possible analysis is based on hardness ratio methods. We have reviewed these $\sim$15~ks \textit{Chandra} observations to obtain 
a measure of the received 2--10~keV flux, and found that it is consistent with a SB origin of only less than $\sim$15 and 30~per cent, respectively, 
of the far-IR emission (see Table~\ref{t2}), assuming that no X-ray contribution from the AGN is detected. \\
Meaningful results have been obtained, instead, for the other two sources, which are definitely among the most extreme ULIRGs. 
IRAS~00182$-$7112, as already mentioned, is shown to harbour a Compton-thick AGN, with flat continuum and prominent Fe~\textsc{xxv} 
K$\alpha$ emission; the reflection component itself is likely to be transmitted through a screen with $N_\rmn{H} \sim 10^{23}$~cm$^{-2}$ 
(Nandra \& Iwasawa 2007). IRAS~08572+3915, instead, has been investigated with all the current X-ray observatories: it exclusively reveals 
a very marginal detection with \textit{Chandra}, but no high-energy emission is seen with \textit{Suzaku} (Teng et al. 2009). CO measurements 
imply for this source a gas column up to $\sim 3 \times 10^{25}$~cm$^{-2}$ (Evans et al. 2002). As a consequence, both these galaxies can 
be considered as a template type~2 quasar in the local Universe. \\
Summarizing, the low number of counts collected from the elusive AGN in our present X-ray ULIRG sample is compatible with the Compton-thickness 
of the putative X-ray absorber, as well as the detection of iron lines with huge equivalent width, resembling the other notable case of Arp~220 
(Iwasawa et al. 2005; Downes \& Eckart 2007). In other words, virtually all ULIRGs with unambiguous mid-IR signatures of powerful, buried 
AGN activity are extinguished in the X-rays, with high incidence of Compton-thick objects and a wide gamut of reflection efficiencies, ranging from 
values expected for a standard type 2 obscuration geometry to much lower ones indicative of full and/or complex covering. 

\section{Discussion}

The X-ray observations analysed in this work provide further insight into the physical and geometrical properties of the gaseous 
material surrounding the AGN and responsible for its obscuration in a ULIRG environment. Assuming the validity of the Compton-thick 
scenario rather than invoking extremely unusual IR to X-ray corrections and/or X-ray reflection efficiencies, the missing AGN detections 
below 10~keV have two main implications which are tightly linked with each other, concerning the covering factor of the X-ray 
absorber and the coupling of the gas and dust components in the outskirts of the compact active source. Also, we discuss our results 
in the context of the AGN/SB connection, searching for possible relations between the excess of AGN obscuration and the strenght of 
the simultaneous burst of star formation. 

\subsection{The nature of the X-ray absorber}

As pointed out earlier, the upper limits that can be placed on the possible AGN reflected emission imply a complete covering of the nuclear 
region. In the X-ray obscured AGN that are classified in the optical as type 2s, the direct optical to soft X-ray radiation blocked by the claimed 
axisymmetric dusty absorber can be scattered into the line of sight after the interactions with material located above the equatorial plane. This 
process strongly affects the shape of the emerging spectrum in the X-rays, and the usual efficiency in terms of the received flux is of the order 
of a few per cent, depending on geometrical effects. Except for the two AGN detections, in all the other sources of our sample we assess a 
reflection efficiency lower than $\sim 2 \times 10^{-3}$, assuming an intrinsic 2--10~keV flux consistent with the results of the mid-IR analysis. 
Thus, regardless of the large and manifold uncertainties involved in the single cases, the X-ray follow-up observations of buried AGN presented 
or reviewed here generally hint at a cocoon-like geometry of the absorber, which is not surprising in extreme and chaotic systems such as 
ULIRGs. \\
Indeed, the detection of so prominent CO gas absorption in objects like IRAS~00182$-$7112 (Spoon et al. 2004) and IRAS~00397$-$1312 
could be related with full covering rather than toroidal obscuration, since this feature is not usually found among type 2 Seyferts (e.g. Lutz et al. 
2004b). As a consequence, the most interesting aspect to investigate concerns the location of the absorber and the properties of its gas and 
dust content. By adopting a typical Galactic extinction curve (e.g. Draine 2003; Nishiyama 2008, 2009) and the standard Galactic dust-to-gas 
ratio (Bohlin, Savage \& Drake 1978), one can easily obtain the relation between the optical depth at 6~$\mu$m and the X-ray column density, 
that is $N_\rmn{H} \simeq 8$--11~$\tau_6 \times 10^{22}$~cm$^{-2}$. Then, even questioning the accuracy of our measure, any reasonable value 
of the \textit{real} $\tau_6$ would not be consistent with a Compton-thick environment with dust and gas components similar to the Galactic diffuse 
interstellar medium. The ensuing point is whether the X-ray column density is physically related to the dust affecting the optical and mid-IR diagnostics. 
There is plenty of observational evidence indicating \textit{anomalous} dust properties in AGN for which different explanations have been proposed. 
For instance, X-ray absorption may occur in the very inner regions, inside the dust sublimation radius (e.g. Granato, Danese \& Franceschini 1997); 
conversely, the formation of large dust grains favoured by high density environments yields a flat extinction curve misleading about the actual dust-to-gas 
ratio (Maiolino, Marconi \& Oliva 2001). \\
In ULIRGs the situation is even more complex: a remarkable example is represented by Mrk~231, a bright ULIRG which is optically classified as a 
\textit{Broad Absorption Line} quasar, and whose fairly unabsorbed \textit{Spitzer}-IRS spectrum suggests a relative AGN contribution to the bolometric 
luminosity of the galaxy of $\sim$30 per cent at least. The 2--10~keV emission of Mrk~231 reveals a reflection component only (Gallagher et al. 2002), 
while the direct AGN continuum shows up at 15--60~keV in the \textit{BeppoSAX} data (Braito et al. 2004). This mismatch between the optical and 
X-ray classification based on the obscuration degree in each band is not infrequent, and suggests a scarce gas/dust coupling around the AGN. 
Moreover, the case of Mrk~231 is indicative of the incidence of outflows: massive gas inflows are needed to sustain nuclear accretion, yet the possible 
link of the X-ray absorbing gas with AGN-driven winds and outflows has far-reaching implications in terms of the AGN feedback and the ULIRG/quasar 
evolutionary pattern. According to this scenario, ULIRGs (or at least a significant fraction of them) are a transitory phase of galactic lifetime preliminary 
to the optically-bright quasar stage, during which the merger-driven inflows provide the gas reservoir for starburst and buried nuclear activity, until the 
AGN feedback disrupts the obscuring shell of dust and gas, quenching the circumnuclear star formation and sweeping the line of sight to the optical 
quasar (Sanders et al. 1988; Hopkins et al. 2008). \\
AGN- or SB-driven galactic winds are known to play a key role in the evolution of galaxies and are found in most ULIRGs (Veilleux, Cecil \& 
Bland-Hawthorn 2005, and references therein). Unambiguous evidence for gas outflows has been recently discovered in the mid-IR also in buried 
AGN candidates. High-resolution \textit{Spitzer}-IRS spectra reveal the presence of high-velocity gas in IRAS~00182$-$7112, giving rise to a broad 
component in the [Ne~\textsc{ii}] and [Ne~\textsc{iii}] lines at 12.81 and 15.56~$\mu$m (Spoon et al. 2009). In both cases the line profile is asymmetric 
due to a blue wing. A similar behaviour is not observed in optical or near-IR forbidden lines, as a result of the large extinction. This blueshift is interpreted 
as the signature of the ongoing breaching of the obscuring medium surrounding the active nucleus. Only two other sources exhibit such an asymmetry 
in neon profiles: our target IRAS~12127$-$1412, and IRAS~13451+1232. The latter is also detected in [Ne~\textsc{v}] at 14.32~$\mu$m, which is a 
straightforward tracer of the AGN radiation field due to the high ionization energy required (97.1~eV). This strongly blueshifted detection indicates a more 
advanced stage of this source in the process of cleaning its line of sight, which is consistent with its type 2 Seyfert classification in the optical. The larger 
part of ULIRGs with [Ne~\textsc{v}] emission show a blueshift, and the higher the neon ionization state the larger the blueshift, proving that the speed 
of the outflow decreases with the distance from the ionizing core (Spoon \& Holt 2009). In some cases the gas kinematics traced by neon lines can be 
associated with the expansion of radio jets. Anyway, out of 25 ULIRGs with either [Ne~\textsc{iii}] or [Ne~\textsc{v}] blueshifted lines, 21 are 
optically-identified AGN.\footnote{Most of these sources are also included in our IR sample.} IRAS~00182$-$7112, IRAS~12127$-$1412 and 
IRAS~01003$-$2238 (neglecting its controversial classification) are among the few exceptions. In this picture, mid-IR buried and X-ray Compton-thick 
AGN are still fully enshrouded, and their feedback is not yet in action. 

\subsection{AGN feedback and SB intensity}

\begin{figure*}
\includegraphics[width=15cm]{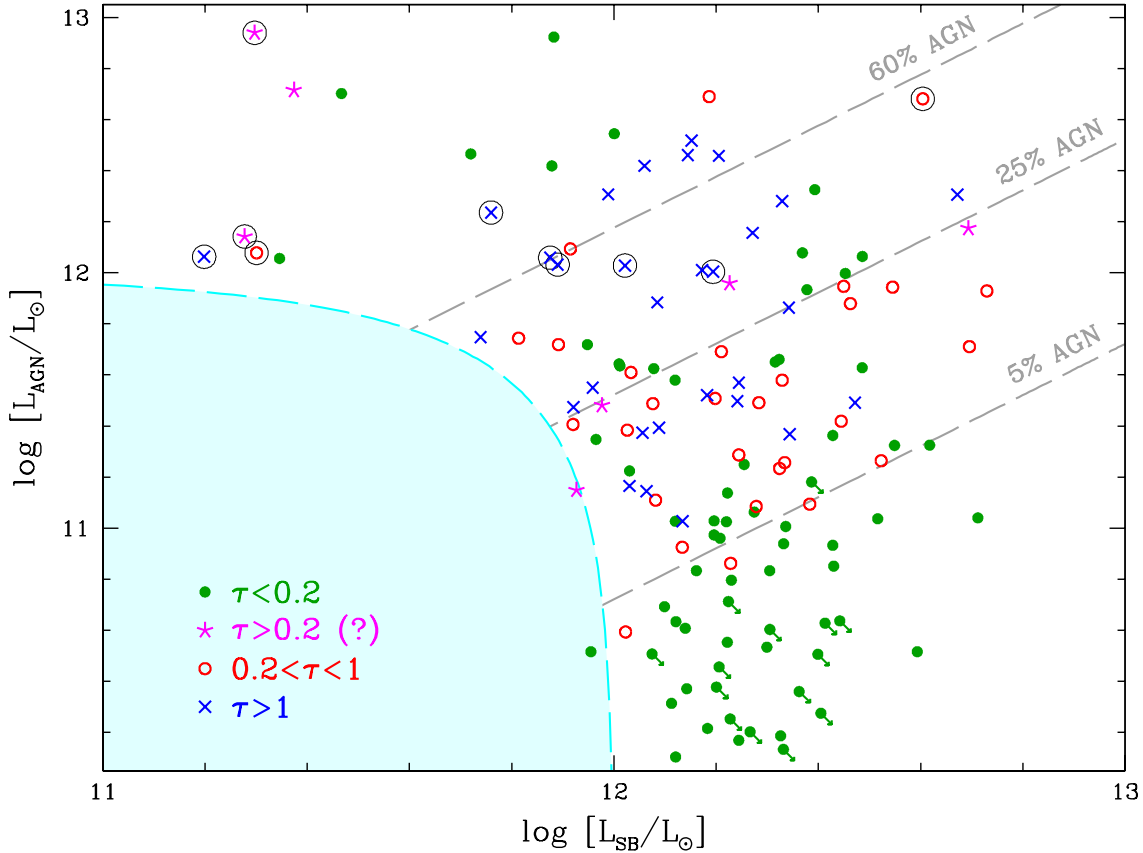}
\caption{AGN versus SB luminosity from our mid-IR analysis. Different colours and symbols are used to distinguish the different 
estimates of the AGN optical depth at 6~$\mu$m. The magenta stars represent those ULIRGs with flat but heavily absorbed 
5--8~$\mu$m spectra, for which the continuum reddening cannot be used to measure the AGN optical depth. Interestingly, the 
largest X-ray reflection efficiencies for buried AGN are found among these objects. The cyan curve describes the boundary 
of our ULIRG sample, $L_\rmn{IR} = L_\rmn{AGN} + L_\rmn{SB} > 10^{12} L_{\sun}$ (all the sources meet this 
condition within the error bars, not shown), while the dashed diagonal lines trace the four regions of the $\alpha_\rmn{bol}$ space 
defined in Nardini et al. (2010). Most of the pure SB/ULIRGs lie outside the plot due to their low upper limit to 
$L_\rmn{AGN}$. Conversely, when the AGN is dominant, the degree of AGN obscuration in the mid-IR seems to be loosely 
related to the intensity of the concurrent star formation. Anyway, since each entry can be somewhat displaced due to the various 
sources of uncertainty (see the text), this plot is only intended for qualitative considerations (consequently, purely statistical error 
bars are omitted for clarity). The encircled objects are the ten ULIRGs presented or reviewed here.}
\label{f5}
\end{figure*}

On the wake of the considerations above, it is worth investigating the possible relation between the AGN mid-IR obscuration and the SB intensity. 
Fierce star formation activity is ubiquitous in AGN both in the local Universe (Schweitzer et al. 2006) and at high redshift (Hatziminaoglou et al. 2010). 
Moreover, far-IR observations suggest that X-ray absorbed AGN suffer a larger contamination from the concurrent star formation than 
unabsorbed ones at the same luminosity and redshift (e.g. Stevens et al. 2005). In order to address this issue, we have plotted in Fig.~\ref{f5} the 
luminosities of the AGN and SB components derived from our mid-IR spectral decomposition, with a code emphasizing the degree of AGN extinction 
at 6~$\mu$m. The relative contribution is indicated by the diagonal lines, which follow the AGN \textit{weight} classification defined in Nardini et al. (2010). 
Before drawing some qualitative considerations, we stress that the exact location of each ULIRG in this plot is affected by several sources of systematic 
uncertainty. In fact, $\alpha_\rmn{bol}$ [here simply given by $(1+L_\rmn{SB}/L_\rmn{AGN})^{-1}$] is an absorption-corrected quantity 
and was calculated as a function of the intrinsic AGN and SB bolometric corrections averaged over the whole sample, but their \textit{actual} 
values for the single objects can be quite different.\footnote{The 1$\sigma$ dispersion inherent in our method is $\sim$0.18~dex (see Nardini et 
al. 2009, 2010 for a review and a comprehensive discussion of our IR analysis).} Also, any deviation from the 
assumed templates and extinction law will modify both $\tau_6$ and $\alpha_\rmn{bol}$, and the difference in the merger stage does not allow 
a uniform interpretation. This notwithstanding, at the higher AGN luminosities (i.e. above $L_\rmn{AGN} \approx 10^{12} L_{\sun}$), 
where all the entries of our X-ray sample lie and the mentioned 
\textit{confusion} effects should be less important, the intensity of the concurrent SB activity seems to be loosely correlated to the AGN mid-IR 
obscuration. Similarly, in terms of growing $\alpha_\rmn{bol}$, when the AGN is still obscured the SB can be very powerful, accounting for roughly 
half of the total luminosity, but the production of stars subsides when the line of sight is clean. \\
At a more speculative level, we focus on the three ULIRGs that have an \textit{anomalous} 5--8~$\mu$m spectral shape with flat but deeply 
absorbed continuum, and are clearly AGN-dominated in the mid-IR. These are the magenta stars in the top left-hand corner of Fig.~\ref{f5}, i.e. 
IRAS~00182$-$7112, IRAS~12127$-$1412 and IRAS~12514+1027. It turns out that all of them are detected as reflected AGN emission in the 
hard X-rays (see Wilman et al. 2003 for the latter object, not included in the present sample). This intriguing case hints at the possibility of 
distinguishing through mid-IR analysis only between a Compton-thickness associated to a toroidal X-ray absorber and a Compton-thickness 
due to a complete AGN covering, coeval with the peak of SB activity. In the former scenario, one can argue that the X-ray reflection signatures 
arise from the inner edge of a slightly tilted torus with little opening angle; this would also imply direct access to a certain fraction of hot dust, possibly 
resulting in a flattening of the continuum in agreement with the spectral trends of quasars around $\sim$5~$\mu$m (see Netzer et al. 2007, and our 
discussion on the AGN template in Nardini et al. 2009). However, further observations are needed to test this conjecture and the chance, if any, to 
infer the X-ray column density and covering factor exclusively from IR diagnostics; this would be relevant to the study of IR galaxies at high redshift 
and to the measure of the contribution to the X-ray background from type 2 quasars. \\
Incidentally, we finally note that by extrapolating the fraction of absorbed AGN in Fig.~\ref{f5} below $\alpha_\rmn{bol}=0.05$, it looks possible that 
almost every ULIRG in the local Universe is the seat of nuclear activity, and that also some of the pure SB/ULIRGs 
(mostly falling outside the plotted region and representing $\sim$30 per cent of the total ULIRG population) harbour an AGN which is both 
intrinsically faint and heavily obscured.\footnote{Such AGN components would be totally missed even with our method, because their contribution 
to the observed mid-IR spectra is too small: in our scheme, these objects are classified as pure SB/ULIRGs.} In this perspective, the clear cut in 
the presence of absorbed sources at $\alpha_\rmn{bol} < 0.05$ is simply due to selection effects: first, when the SB dominates at even 
5--8~$\mu$m, the AGN optical depth can be understimated, and some of the green dots would turn into red circles with just a tiny displacement; 
second, some of the non-detections would be shifted upwards, entering the plotted portion of the $\alpha_\rmn{bol} < 0.05$ region as either 
red circles or blue crosses. 

\section{Conclusions}

Our recent mid-IR diagnostic campaign on local ULIRGs has revealed the presence in a sizable fraction of these sources of powerful 
AGN components, providing a significant contribution to the total energy output but missed at optical wavelengths because of their heavy 
obscuration. The mid-IR spectra of these intriguing sources provide strong evidence of buried nuclear activity, in the shape of a steep and 
intense continuum, dramatic suppression of PAH features and various absorption. In order to pursue a more exhaustive study of the extreme 
environment surrounding the AGN component inside this subclass of ULIRGs, we have presented the X-ray analysis of a representative 
sample consisting of ten objects of this kind, and including our new follow-up X-ray observations obtained with \textit{Chandra} and 
\textit{XMM-Newton} of IRAS~00397$-$1312, IRAS~01003$-$2238, IRAS~01298$-$0744 and IRAS~12127$-$1412. Only in 
IRAS~12127$-$1412 the presence of the AGN is clearly established in the X-rays, while IRAS~01298$-$0744 is not detected at all. The 
faint X-ray emission of IRAS~00397$-$1312 and IRAS~01003$-$2238 is consistent with a pure SB contribution, in terms of both spectral 
shape and observed flux. The other six sources in our X-ray sample, two of which have not been published to date, fully validate this general 
trend, with only another safe AGN detection in IRAS~00182$-$7112. \\
The main implications of this work are the following: 1) Assuming a typical dust extinction law, the dust-to-gas ratio is required to be much lower 
than the Galactic standard in order to support the Compton-thick interpretation for the unsuccessful AGN detections. Of course, this is not due to 
a large-scale dust deficiency, since ULIRGs are, by definition, dust-rich systems. Anyway, absorption features such as that of CO at 4.65~$\mu$m 
observed in IRAS~00397$-$1312 hint at large amounts of gas concentrated nearby a compact active core. A possible imbalance between the dust 
and gas components involving only the nuclear regions (i.e. within a region comparable in size with the dust sublimation radius) may be related to 
massive gaseous inflows/outflows during the most active phases of AGN accretion. 2) Whenever the AGN is not detected but the amplitude of a 
possible reflection component can be constrained, its upper limit is $>500$ times lower than the expected intrinsic 2--10~keV AGN flux. This is an 
order of magnitude lower than the usual reflection efficiency observed in type 2 active galaxies, suggesting an almost complete AGN covering in these 
ULIRGs. It is unlikely that the standard toroidal geometry of the dust absorber predicted for isolated and morphologically quiet AGN will hold 
systematically also for much more complex systems like ULIRGs, whose geometrical structure can be poorly retraced. Again, a complete covering is 
compatible with a spherical shell of gas infalling on to the accreting SMBH or, more conceivable in advanced merger stages, with the dust-free base 
of the winds invoked in the feedback scenario to clean the line of sight to the AGN and, possibly, to quench star formation. \\
The elusive AGN identified through their mid-IR spectral properties, the most notable of which have been discussed in this work, are likely going 
through the ultimate phase of buried activity, and are therefore the ideal sources to search for detectable signatures of the impending transition from 
ULIRGs to optically bright quasars. Even if the quality of the single X-ray spectra is very modest, the availability of X-ray constraints is crucial to derive 
a comprehensive multiwavelength picture. Extending the number of intrinsically luminous but heavily obscured AGN for which it is possible to infer the 
amount and distribution of gas in the circumnuclear environment and to compare such information with the optical to mid-IR dust extinction and absorption 
features and the intensity of the concurrent star formation may shed light on the ULIRG/quasar evolutionary sequence and the AGN feedback mechanisms 
affecting the properties of the host galaxy. Moreover, if the tentative correlation between mid-IR spectral properties and X-ray column density and covering 
factor is confirmed and clearly established, a simple \textit{qualitative} inspection of the mid-IR spectra of the high-redshift counterparts of local ULIRGs will 
unveil Compton-thick AGN candidates and enable a measure of their contribution to the X-ray background. 

\section*{Acknowledgments}

This research has been partially supported by NASA grants NNX09AT10G and GO0-11017X. 
We thank the anonymous referee for his/her constructive comments which improved our work. 


\label{lastpage}


\begin{thebibliography}{}

\bibitem[\protect\citeauthoryear{Alexander et 
al.}{2003}]{2003AJ....126..539A} Alexander D.~M., et al., 2003, AJ, 126, 
539
\bibitem[\protect\citeauthoryear{Alexander et 
al.}{2005}]{2005ApJ...632..736A} Alexander D.~M., Bauer F.~E., Chapman 
S.~C., Smail I., Blain A.~W., Brandt W.~N., Ivison R.~J., 2005, ApJ, 632, 
736
\bibitem[\protect\citeauthoryear{Allen et al.}{1991}]{1991MNRAS.248..528A} 
Allen D.~A., Norris R.~P., Meadows V.~S., Roche P.~F., 1991, MNRAS, 248, 
528
\bibitem[\protect\citeauthoryear{Antonucci}{1993}]{1993ARA&A..31..473A} 
Antonucci R., 1993, ARA\&A, 31, 473
\bibitem[\protect\citeauthoryear{Bauer et al.}{2010}]{2010ApJ...710..212B} 
Bauer F.~E., Yan L., Sajina A., Alexander D.~M., 2010, ApJ, 710, 212
\bibitem[\protect\citeauthoryear{Bohlin, Savage, 
\& Drake}{1978}]{1978ApJ...224..132B} Bohlin R.~C., Savage B.~D., Drake J.~F., 1978, ApJ, 224, 132
\bibitem[\protect\citeauthoryear{Braito et 
al.}{2004}]{2004A&A...420...79B} Braito V., et al., 2004, A\&A, 420, 79
\bibitem[\protect\citeauthoryear{Brandl et al.}{2006}]{2006ApJ...653.1129B}
Brandl B.~R., et al., 2006, ApJ, 653, 1129
\bibitem[\protect\citeauthoryear{Daddi et al.}{2007}]{2007ApJ...670..173D} 
Daddi E., et al., 2007, ApJ, 670, 173
\bibitem[\protect\citeauthoryear{Downes
\& Eckart}{2007}]{2007A&A...468L..57D} Downes D., Eckart A., 2007, A\&A, 468, L57
\bibitem[\protect\citeauthoryear{Draine}{2003}]{2003ARA&A..41..241D} Draine B.~T., 2003, ARA\&A, 41, 241
\bibitem[\protect\citeauthoryear{Elvis et al.}{1994}]{1994ApJS...95....1E} 
Elvis M., et al., 1994, ApJS, 95, 1
\bibitem[\protect\citeauthoryear{Evans et al.}{2002}]{2002ApJ...580..749E} 
Evans A.~S., Mazzarella J.~M., Surace J.~A., Sanders D.~B., 2002, ApJ, 580, 749
\bibitem[\protect\citeauthoryear{Fabian 
\& Iwasawa}{1999}]{1999MNRAS.303L..34F} Fabian A.~C., Iwasawa K., 1999, MNRAS, 303, L34
\bibitem[\protect\citeauthoryear{Fiore et al.}{2008}]{2008ApJ...672...94F} 
Fiore F., et al., 2008, ApJ, 672, 94
\bibitem[\protect\citeauthoryear{Franceschini et 
al.}{2003}]{2003MNRAS.343.1181F} Franceschini A., et al., 2003, MNRAS, 343, 
1181 
\bibitem[\protect\citeauthoryear{Gallagher et 
al.}{2002}]{2002ApJ...569..655G} Gallagher S.~C., Brandt W.~N., Chartas G., 
Garmire G.~P., Sambruna R.~M., 2002, ApJ, 569, 655
\bibitem[\protect\citeauthoryear{Genzel et al.}{1998}]{1998ApJ...498..579G} 
Genzel R., et al., 1998, ApJ, 498, 579
\bibitem[\protect\citeauthoryear{Gilli, Comastri, 
\& Hasinger}{2007}]{2007A&A...463...79G} Gilli R., Comastri A., Hasinger G., 2007, A\&A, 463, 79
\bibitem[\protect\citeauthoryear{Goulding et 
al.}{2011}]{2011MNRAS.411.1231G} Goulding A.~D., Alexander D.~M., Mullaney 
J.~R., Gelbord J.~M., Hickox R.~C., Ward M., Watson M.~G., 2011, MNRAS, 
411, 1231
\bibitem[\protect\citeauthoryear{Granato, Danese, 
\& Franceschini}{1997}]{1997ApJ...486..147G} Granato G.~L., Danese L., Franceschini A., 1997, ApJ, 486, 147
\bibitem[\protect\citeauthoryear{Hatziminaoglou et 
al.}{2010}]{2010A&A...518L..33H} Hatziminaoglou E., et al., 2010, A\&A, 518, L33
\bibitem[\protect\citeauthoryear{Helou, Soifer, 
\& Rowan-Robinson}{1985}]{1985ApJ...298L...7H} Helou G., Soifer B.~T., Rowan-Robinson M., 1985, ApJ, 298, L7
\bibitem[\protect\citeauthoryear{Hinshaw et 
al.}{2009}]{2009ApJS..180..225H} Hinshaw G., et al., 2009, ApJS, 180, 225
\bibitem[\protect\citeauthoryear{Hopkins et 
al.}{2008}]{2008ApJS..175..356H} Hopkins P.~F., Hernquist L., Cox T.~J., 
Kere{\v s} D., 2008, ApJS, 175, 356
\bibitem[\protect\citeauthoryear{Houck et al.}{2004}]{2004ApJS..154...18H}
Houck J.~R., et al., 2004, ApJS, 154, 18
\bibitem[\protect\citeauthoryear{Kartaltepe et 
al.}{2010}]{2010ApJ...709..572K} Kartaltepe J.~S., et al., 2010, ApJ, 709, 
572 
\bibitem[\protect\citeauthoryear{Kim \&
Sanders}{1998}]{1998ApJS..119...41K} Kim D.-C., Sanders D.~B., 1998, ApJS,
119, 41
\bibitem[\protect\citeauthoryear{Imanishi}{2009}]{2009ApJ...694..751I} 
Imanishi M., 2009, ApJ, 694, 751
\bibitem[\protect\citeauthoryear{Imanishi et 
al.}{2008}]{2008PASJ...60S.489I} Imanishi M., Nakagawa T., Ohyama Y., 
Shirahata M., Wada T., Onaka T., Oi N., 2008, PASJ, 60, 489
\bibitem[\protect\citeauthoryear{Imanishi et 
al.}{2010}]{2010ApJ...721.1233I} Imanishi M., Nakagawa T., Shirahata M., 
Ohyama Y., Onaka T., 2010, ApJ, 721, 1233
\bibitem[\protect\citeauthoryear{Iwasawa et
al.}{2005}]{2005MNRAS.357..565I} Iwasawa K., Sanders D.~B., Evans A.~S.,
Trentham N., Miniutti G., Spoon H.~W.~W., 2005, MNRAS, 357, 565
\bibitem[\protect\citeauthoryear{Laurent et 
al.}{2000}]{2000A&A...359..887L} Laurent O., Mirabel I.~F., Charmandaris V., Gallais P., Madden S.~C., Sauvage M., Vigroux L., Cesarsky C., 2000, A\&A, 359, 887
\bibitem[\protect\citeauthoryear{Lusso et 
al.}{2010}]{2010A&A...512A..34L} Lusso E., et al., 2010, A\&A, 512, A34
\bibitem[\protect\citeauthoryear{Lutz et 
al.}{2004}]{2004A&A...418..465L} Lutz D., Maiolino R., Spoon H.~W.~W., Moorwood A.~F.~M., 2004b, A\&A, 418, 465
\bibitem[\protect\citeauthoryear{Lutz et 
al.}{2004}]{2004A&A...426L...5L} Lutz D., Sturm E., Genzel R., Spoon H.~W.~W., Stacey G.~J., 2004a, A\&A, 426, L5
\bibitem[\protect\citeauthoryear{Magdziarz 
\& Zdziarski}{1995}]{1995MNRAS.273..837M} Magdziarz P., Zdziarski A.~A., 1995, MNRAS, 273, 837
\bibitem[\protect\citeauthoryear{Maiolino, Marconi, 
\& Oliva}{2001}]{2001A&A...365...37M} Maiolino R., Marconi A., Oliva E., 2001, A\&A, 365, 37
\bibitem[\protect\citeauthoryear{Matt, Brandt, 
\& Fabian}{1996}]{1996MNRAS.280..823M} Matt G., Brandt W.~N., Fabian A.~C., 1996, MNRAS, 280, 823
\bibitem[\protect\citeauthoryear{Nandra 
\& Iwasawa}{2007}]{2007MNRAS.382L...1N} Nandra K., Iwasawa K., 2007, MNRAS, 382, L1
\bibitem[\protect\citeauthoryear{Nardini et 
al.}{2008}]{2008MNRAS.385L.130N} Nardini E., Risaliti G., Salvati M., Sani 
E., Imanishi M., Marconi A., Maiolino R., 2008, MNRAS, 385, L130
\bibitem[\protect\citeauthoryear{Nardini et 
al.}{2009}]{2009MNRAS.399.1373N} Nardini E., Risaliti G., Salvati M., Sani 
E., Watabe Y., Marconi A., Maiolino R., 2009, MNRAS, 399, 1373
\bibitem[\protect\citeauthoryear{Nardini et 
al.}{2010}]{2010MNRAS.405.2505N} Nardini E., Risaliti G., Watabe Y., 
Salvati M., Sani E., 2010, MNRAS, 405, 2505
\bibitem[\protect\citeauthoryear{Netzer et al.}{2007}]{2007ApJ...666..806N}
Netzer H., et al., 2007, ApJ, 666, 806
\bibitem[\protect\citeauthoryear{Nishiyama et 
al.}{2008}]{2008ApJ...680.1174N} Nishiyama S., Nagata T., Tamura M., 
Kandori R., Hatano H., Sato S., Sugitani K., 2008, ApJ, 680, 1174
\bibitem[\protect\citeauthoryear{Nishiyama et 
al.}{2009}]{2009ApJ...696.1407N} Nishiyama S., Tamura M., Hatano H., Kato 
D., Tanab{\'e} T., Sugitani K., Nagata T., 2009, ApJ, 696, 1407
\bibitem[\protect\citeauthoryear{Ptak et al.}{2003}]{2003ApJ...592..782P} 
Ptak A., Heckman T., Levenson N.~A., Weaver K., Strickland D., 2003, ApJ, 
592, 782
\bibitem[\protect\citeauthoryear{Ranalli, Comastri, 
\& Setti}{2003}]{2003A&A...399...39R} Ranalli P., Comastri A., Setti G., 2003, A\&A, 399, 39
\bibitem[\protect\citeauthoryear{Risaliti et 
al.}{2000}]{2000A&A...357...13R} Risaliti G., Gilli R., Maiolino R., Salvati M., 2000, A\&A, 357, 13
\bibitem[\protect\citeauthoryear{Sanders \&
Mirabel}{1996}]{1996ARA&A..34..749S} Sanders D.~B., Mirabel I.~F., 1996,
ARA\&A, 34, 749
\bibitem[\protect\citeauthoryear{Sanders et
al.}{1988}]{1988ApJ...328L..35S} Sanders D.~B., Soifer B.~T., Elias J.~H.,
Neugebauer G., Matthews K., 1988, ApJ, 328, L35
\bibitem[\protect\citeauthoryear{Sanders et 
al.}{2003}]{2003AJ....126.1607S} Sanders D.~B., Mazzarella J.~M., Kim 
D.-C., Surace J.~A., Soifer B.~T., 2003, AJ, 126, 1607
\bibitem[\protect\citeauthoryear{Sani et al.}{2008}]{2008ApJ...675...96S} 
Sani E., et al., 2008, ApJ, 675, 96
\bibitem[\protect\citeauthoryear{Schweitzer et
al.}{2006}]{2006ApJ...649...79S} Schweitzer M., et al., 2006, ApJ, 649, 79
\bibitem[\protect\citeauthoryear{Spoon 
\& Holt}{2009}]{2009ApJ...702L..42S} Spoon H.~W.~W., Holt J., 2009, ApJ, 702, L42 
\bibitem[\protect\citeauthoryear{Spoon et al.}{2004}]{2004ApJS..154..184S} 
Spoon H.~W.~W., et al., 2004, ApJS, 154, 184
\bibitem[\protect\citeauthoryear{Spoon et al.}{2009}]{2009ApJ...693.1223S} 
Spoon H.~W.~W., Armus L., Marshall J.~A., Bernard-Salas J., Farrah D., 
Charmandaris V., Kent B.~R., 2009, ApJ, 693, 1223 
\bibitem[\protect\citeauthoryear{Stevens et 
al.}{2005}]{2005MNRAS.360..610S} Stevens J.~A., Page M.~J., Ivison R.~J., 
Carrera F.~J., Mittaz J.~P.~D., Smail I., McHardy I.~M., 2005, MNRAS, 360, 
610
\bibitem[\protect\citeauthoryear{Teng 
\& Veilleux}{2010}]{2010ApJ...725.1848T} Teng S.~H., Veilleux S., 2010, ApJ, 725, 1848
\bibitem[\protect\citeauthoryear{Teng et al.}{2005}]{2005ApJ...633..664T} 
Teng S.~H., Wilson A.~S., Veilleux S., Young A.~J., Sanders D.~B., Nagar 
N.~M., 2005, ApJ, 633, 664
\bibitem[\protect\citeauthoryear{Teng et al.}{2009}]{2009ApJ...691..261T} 
Teng S.~H., et al., 2009, ApJ, 691, 261
\bibitem[\protect\citeauthoryear{Treister et 
al.}{2009}]{2009ApJ...706..535T} Treister E., et al., 2009, ApJ, 706, 535
\bibitem[\protect\citeauthoryear{Veilleux, Kim, \&
Sanders}{1999}]{1999ApJ...522..113V} Veilleux S., Kim D.-C., Sanders D.~B.,
1999, ApJ, 522, 113
\bibitem[\protect\citeauthoryear{Veilleux, Kim, 
\& Sanders}{2002}]{2002ApJS..143..315V} Veilleux S., Kim D.-C., Sanders D.~B., 2002, ApJS, 143, 315
\bibitem[\protect\citeauthoryear{Veilleux, Cecil, 
\& Bland-Hawthorn}{2005}]{2005ARA&A..43..769V} Veilleux S., Cecil G., Bland-Hawthorn J., 2005, ARA\&A, 43, 769
\bibitem[\protect\citeauthoryear{Werner et al.}{2004}]{2004ApJS..154....1W}
Werner M.~W., et al., 2004, ApJS, 154, 1
\bibitem[\protect\citeauthoryear{Wilman et al.}{2003}]{2003MNRAS.338L..19W} 
Wilman R.~J., Fabian A.~C., Crawford C.~S., Cutri R.~M., 2003, MNRAS, 338, L19
\bibitem[\protect\citeauthoryear{Yuan, Kewley, 
\& Sanders}{2010}]{2010ApJ...709..884Y} Yuan T.-T., Kewley L.~J., Sanders D.~B., 2010, ApJ, 709, 884

\end{thebibliography}
\end{document}